\documentclass[conference]{IEEEtran}
\IEEEoverridecommandlockouts

\usepackage{amsmath,amssymb,amsfonts}   %
\usepackage{textcomp}                   %

\usepackage{graphicx}                   %
\usepackage{xcolor}                     %

\usepackage{array}                      %
\newcolumntype{H}{>{\setbox0=\hbox\bgroup}c<{\egroup}@{}} %
\usepackage{booktabs}                   %
\usepackage{siunitx}                    %
  \DeclareSIUnit{\dBm}{dBm}
  \DeclareSIUnit{\dBi}{dBi}
  \DeclareSIUnit{\dBsm}{dBsm}

\usepackage{soul}                       %

\usepackage[backend=biber,
            language=english,
            autolang=other,
            style=ieee]{biblatex}       %
\usepackage[dvipsnames,svgnames,table]{xcolor} %
\usepackage[hidelinks]{hyperref}               %
\usepackage[capitalize]{cleveref}                          %

\usepackage{algorithmic}                %

\usepackage{pgfplots}                   %
  \pgfplotsset{compat=newest}
\usepgfplotslibrary{groupplots,dateplot,smithchart} %
\usepackage{pgfplotstable}              %
  \pgfplotsset{compat=1.18}              %

\usetikzlibrary{%
  patterns,          %
  shapes.arrows,     %
  spy,               %
  external,          %
  calc,              %
  shadings,          %
  shadows.blur,      %
  decorations.pathreplacing %
}
\usepackage{circuitikz}

\definecolor{mplblue}{RGB}{31,119,180}
\definecolor{mplorange}{RGB}{255,127,14}
\definecolor{mplgreen}{RGB}{44,160,44}
\definecolor{mplred}{RGB}{214,39,40}
\definecolor{mplpurple}{RGB}{148,103,189}
\definecolor{mplbrown}{RGB}{140,86,75}
\definecolor{mplpink}{RGB}{227,119,194}
\definecolor{mplgray}{RGB}{127,127,127}
\definecolor{mployellow}{RGB}{188,189,34}
\definecolor{mplcyan}{RGB}{23,190,207}

\pgfplotsset{
  cycle list={
    {mplblue,solid},
    {mplorange,solid},
    {mplgreen,solid},
    {mplred,solid},
    {mplpurple,solid},
    {mplbrown,solid},
    {mplpink,solid},
    {mplgray,solid},
    {mployellow,solid},
    {mplcyan,solid},
  },
  every axis plot/.append style={very thick, line join=round}
}

\usepackage[xindy,nonumberlist]{glossaries} %
  \makenoidxglossaries            %
  
\newacronym{2d}{2D}{two-dimensional}
\newacronym{3d}{3D}{three-dimensional}
\newacronym{3gpp}{3GPP}{3rd generation partnership project}
\newacronym{3msk}{3MSK}{3-level minimum-shift-keying}
\newacronym{4g}{4G}{fourth-generation}
\newacronym{5g}{5G}{fifth-generation}
\newacronym{6dof}{6DoF}{six degrees of freedom}
\newacronym{6g}{6G}{sixth-generation}
\newacronym{abp}{ABP}{authentication by personalisation}
\newacronym{ac}{AC}{alternating current}
\newacronym{aci}{ACI}{adjacent channel interference}
\newacronym{aclr}{ACLR}{adjacent channel leakage ratio}
\newacronym{acpr}{ACPR}{adjacent channel power ratio}
\newacronym{adc}{ADC}{analog-to-digital converter}
\newacronym{adr}{ADR}{adaptive data rate}
\newacronym{aec}{AEC}{aluminum electrolytic capacitor}
\newacronym{aes}{AES}{Advanced Encryption Standard}
\newacronym{afe}{AFE}{analog front end}
\newacronym{ag}{AG}{array gain}
\newacronym{agc}{AGC}{automatic gain controller}
\newacronym{agps}{A-GPS}{Assisted Global Positioning System} %
\newacronym{ai}{AI}{artificial intelligence}
\newacronym{alk}{ALK}{Alkaline}
\newacronym{am}{AM}{amplitude modulation}
\newacronym{amam}{AM/AM}{amplitude modulation to amplitude modulation}
\newacronym{amc}{AMC}{automatic modulation classification}
\newacronym{amp}{AMP}{approximate message passing}
\newacronym{ampm}{AM/PM}{amplitude modulation to phase modulation}
\newacronym{aoa}{AOA}{angle-of-arrival}
\newacronym{aod}{AOD}{angle-of-departure}
\newacronym{ap}{AP}{access point}
\newacronym{api}{API}{application program interface}
\newacronym{apt}{APT}{acoustic power transfer}
\newacronym{apu}{APU}{access point unit}
\newacronym{ar}{AR}{augmented reality}
\newacronym{arima}{ARIMA}{Auto-Regressive Integrated Moving Average}
\newacronym{arp}{ARP}{Antenna Reference Point}
\newacronym{asic}{ASIC}{application specific integrated circuit}
\newacronym{ask}{ASK}{amplitude-shift keying}
\newacronym{at}{AT}{ATtention}
\newacronym{auv}{AUV}{autonomous underwater vehicle}
\newacronym{awg}{AWG}{arbitrary waveform generator}
\newacronym{awgn}{AWGN}{additive white Gaussian noise}
\newacronym{baw}{BAW}{bulk acoustic wave}
\newacronym{bb}{BB}{base-band}
\newacronym{bc}{BC}{backscatter communication}
\newacronym{bcjr}{BCJR}{Bahl-Cocke-Jelinek-Raviv}
\newacronym{bd}{BD}{backscatter device}
\newacronym{be}{BE}{Belgium}
\newacronym{ber}{BER}{bit error rate}
\newacronym{bf}{BF}{beamforming}
\newacronym{bga}{BGA}{ball grid array}
\newacronym{bldc}{BLDC}{brushless DC}
\newacronym{bler}{BLER}{block error rate}
\newacronym{bod}{BOD}{Brown-Out Detection}
\newacronym{bom}{BOM}{bill of materials}
\newacronym{bpsk}{BPSK}{binary phase-shift keying}
\newacronym{brp}{BRP}{beam refinement process}
\newacronym{bs}{BS}{base station}
\newacronym{bu}{BU}{booster unit}
\newacronym{bw}{BW}{bandwidth}
\newacronym{ca}{CA}{carrier emitter}
\newacronym{cad}{CAD}{channel activity detection}
\newacronym{cars}{CARS}{calibration reference signal}
\newacronym{cbm}{CBM}{condition based maintenance}
\newacronym{cc}{CC}{constant current}
\newacronym{ccdf}{CCDF}{complementary cumulative distribution function}
\newacronym{ccnn}{CCNN}{circular convolutional neural network}
\newacronym{ccs}{CCS}{correlative channel sounder}
\newacronym{cdf}{CDF}{cumulative distribution function}
\newacronym{cdma}{CDMA}{code division-multiple access}
\newacronym{cdrx}{CDRX}{connected mode DRX}
\newacronym{ce}{CE}{coverage enhancement}
\newacronym{ced}{CED}{cumulative energy density}
\newacronym{ceofdm}{CE-OFDM}{constant-envelope OFDM}
\newacronym{cf}{CF}{cell-free}
\newacronym{cfmmimo}{CF-mMIMO}{cell-free massive MIMO}
\newacronym{cfo}{CFO}{carrier frequency offset}
\newacronym{cir}{CIR}{channel impulse response}
\newacronym{cla}{CLA}{closed-loop approach}
\newacronym{clk}{CLK}{clock}
\newacronym{cmos}{CMOS}{complementary metal oxide semiconductor}
\newacronym{cnn}{CNN}{convolutional neural network}
\newacronym{co}{CO}{combinatorial optimization}
\newacronym{cordic}{CORDIC}{coordinate rotation digital computer}
\newacronym{cost}{COST}{commercial off-the-shelf}
\newacronym{cots}{COTS}{commercial off-the-shelf}
\newacronym{cp}{CP}{cyclic prefix}
\newacronym{cpe}{CPE}{common phase error}
\newacronym{cpfsk}{CPFSK}{continuous phase frequency shift keying}
\newacronym{cpm}{CPM}{continuous phase modulation}
\newacronym{cpt}{CPT}{capacitive power transfer}
\newacronym{cpu}{CPU}{central-processing unit}
\newacronym{cpw}{CPW}{coplanar waveguide}
\newacronym{cqi}{CQI}{channel quality indicator}
\newacronym{cr}{CR}{coding rate}
\newacronym{crc}{CRC}{cyclic redundancy check}
\newacronym{crlb}{CRLB}{Cram\'er-Rao lower bound}
\newacronym{crs}{CRS}{cell reference signal}
\newacronym{cs}{CS}{compressed sensing}
\newacronym{cse}{CSE}{channel state estimation}
\newacronym{csi}{CSI}{channel state information}
\newacronym{csp}{CSP}{contact service point}
\newacronym{css}{CSS}{chirp spread spectrum}
\newacronym{cu}{CU}{central unit}
\newacronym{cv}{CV}{constant voltage}
\newacronym{cw}{CW}{continuous wave}
\newacronym{d2d}{D2D}{device-to-device}
\newacronym{dab}{DAB}{Digital Audio Broadcasting}
\newacronym{dac}{DAC}{digital-to-analog converter}
\newacronym{daq}{DAQ}{data acquisition system}
\newacronym{das}{DAS}{distributed antenna systems}
\newacronym{dbpsk}{DBPSK}{Differential Binary Phase Shift Keying}
\newacronym{dc}{DC}{direct current}
\newacronym{dcc}{DCC}{dynamic cooperation clustering}
\newacronym{ddc}{DDC}{digital down conversion}
\newacronym{de}{DE}{drain efficiency}
\newacronym{dft}{DFT}{discrete Fourier transform}
\newacronym{dftsofdm}{DFT-s-OFDM}{discrete Fourier Transform spread OFDM}
\newacronym{dftsofdmfdss}{DFT-s-OFDM-FDSS}{DFT-s-OFDM with frequency-domain spectral shaping}
\newacronym{dftsofdmfdssse}{DFT-s-OFDM-FDSS-SE}{DFT-s-OFDM with FDSS and spectral extension}
\newacronym{dl}{DL}{downlink}
\newacronym{dlc}{DLC}{Distributed Laser Charging}
\newacronym{dli}{DLI}{direct link interference}
\newacronym{dlt}{DLT}{Distributed Ledger Technology}
\newacronym{dm}{DM}{diffuse multipath}
\newacronym{dma}{DMA}{Direct Memory Access}
\newacronym{dmac}{DMAC}{Direct Memory Access Controller}
\newacronym{dmc}{DMC}{diffuse multipath component}
\newacronym{dmimo}{D-MIMO}{distributed MIMO}
\newacronym{dnn}{DNN}{deep neural network}
\newacronym{doa}{DOA}{direction-of-arrival}
\newacronym{dp}{DP}{differencial privacy}
\newacronym{dpd}{DPD}{digital pre-distortion}
\newacronym{dpdk}{DPDK}{Data Plane Development Kit}
\newacronym{dram}{DRAM}{dynamic random-access memory}
\newacronym{drcs}{$\Delta$RCS}{differential-radar cross section}
\newacronym{drx}{DRX}{Discontinuous Reception Mode}
\newacronym{dsb}{DSB}{double-sideband}
\newacronym{dsl}{DSL}{digital subscriber line}
\newacronym{dsp}{DSP}{digital signal processing}
\newacronym{dss}{DSS}{Dataset Storage Standard}
\newacronym{duc}{DUC}{digital up-converter}
\newacronym{dvb}{DVB}{Digital Video Broadcasting}
\newacronym{e2e}{E2E}{end-to-end}
\newacronym{easa}{EASA}{European Union Aviation Safety Agency}
\newacronym{ebg}{EBG}{electromagnetic bandgap}
\newacronym{ebw}{EBW}{excess bandwidth}
\newacronym{ec}{EC}{European Commission}
\newacronym{ecc}{ECC}{elliptic curve cryptography}
\newacronym{ecdf}{eCDF}{empirical cumulative distribution function}
\newacronym{ecdlp}{ECDLP}{elliptic curve discrete logarithm problem}
\newacronym{ecsp}{ECSP}{edge computing service point}
\newacronym{edlc}{EDLC}{electrostatic double-layer capacitors}
\newacronym{edrx}{eDRX}{Extended Discontinuous Reception Mode}
\newacronym{ee}{EE}{energy efficiency}
\newacronym{egprs}{EGPRS}{Enhanced Data Rates for GSM Evolution}
\newacronym{eh}{EH}{energy harvesting}
\newacronym{eirp}{EIRP}{equivalent isotropically radiated power}
\newacronym{em}{EM}{electromagnetic}
\newacronym{embb}{eMBB}{enhanced Mobile Broadband}
\newacronym{en}{EN}{energy neutral}
\newacronym{end}{END}{energy-neutral device}
\newacronym{enob}{ENOB}{effective number of bits}
\newacronym{eol}{EoL}{end of life}
\newacronym{ep}{EP}{energy profiler}
\newacronym{epd}{EPD}{electronic paper display}
\newacronym{epu}{EPU}{edge processing unit}
\newacronym{er}{ER}{Energy Receiver}
\newacronym{erp}{ERP}{effective radiation power}
\newacronym[plural=ESCs,firstplural=electronic speed controllers (ESCs)]{esc}{ESC}{electronic speed control}
\newacronym{esd}{ESD}{electrostatic discharge}
\newacronym{esl}{ESL}{electronic shelf label}
\newacronym{esr}{ESR}{equivalent series resistance}
\newacronym{et}{ET}{Energy Transmitter}
\newacronym{etsi}{ETSI}{European Telecommunications Standards Institute}
\newacronym{evd}{EVD}{eigenvalue decomposition}
\newacronym{evm}{EVM}{Error Vector Magnitude}
\newacronym{ewlb}{eWLB}{Embedded Wafer Level Ball Grid Array}
\newacronym{fa}{FA}{federation anchor}
\newacronym{fair}{FAIR}{Findability, Accessibility, Interoperability, and Reuse of digital assets}
\newacronym{fcf}{FCF}{frequency correlation function}
\newacronym{fd}{FD}{front-haul distance}
\newacronym{fdd}{FDD}{frequency-division duplexing}
\newacronym{fde}{FDE}{frequency domain equalizer}
\newacronym{fdm}{FDM}{frequency-division multiplexing}
\newacronym{fdma}{FDMA}{frequency division multiple access}
\newacronym{fdss}{FDSS}{frequency-domain spectral shaping}
\newacronym{fem}{FEM}{finite element analysis}
\newacronym{fembb}{feMBB}{further enhanced mobile broadband}
\newacronym{fft}{FFT}{fast Fourier transform}
\newacronym{fh}{FH}{fronthaul}
\newacronym{fhss}{FHSS}{frequency hopping spread spectrum}
\newacronym{fifo}{FIFO}{First In, First Out}
\newacronym{fim}{FIM}{Fisher information matrix}
\newacronym{fits}{FITS}{Flexible Image Transport System}
\newacronym{fl}{FL}{federated learning}
\newacronym{fom}{FoM}{figure of merit}
\newacronym{fov}{FOV}{field of view}
\newacronym{fpga}{FPGA}{field-programmable gate array}
\newacronym{fr1}{FR1}{frequency range 1}
\newacronym{fr2}{FR2}{frequency range 2}
\newacronym{fram}{FRAM}{Ferroelectric Random Access Memory}
\newacronym{fsk}{FSK}{frequency shift keying}
\newacronym{fspl}{FSPL}{free space path loss}
\newacronym{fss}{FSS}{frequency selective surface}
\newacronym{gb}{GB}{grant-based}
\newacronym{gdpr}{GDPR}{general data protection regulation}
\newacronym{gf}{GF}{grant-free}
\newacronym{gmsk}{GMSK}{Gaussian minimum-shift keying}
\newacronym{gnb}{gNB}{Next Generation Node B}
\newacronym{gni}{GNI}{gross national income}
\newacronym{gnn}{GNN}{graph neural network}
\newacronym{gnss}{GNSS}{global navigation satellite system}
\newacronym{gpclk}{GPCLK}{general purpose clock}
\newacronym{gpio}{GPIO}{General-Purpose Input/Output}
\newacronym{gpl}{GPL}{GNU General Public License}
\newacronym{gprs}{GPRS}{General Packet Radio Services}
\newacronym{gps}{GPS}{Global Positioning System}
\newacronym{gpu}{GPU}{graphical processing unit}
\newacronym{grc}{GRC}{GNU Radio Companion}
\newacronym{gscm}{GSCM}{geometry‐based stochastic model}
\newacronym{gsm}{GSM}{Global System for Mobile Communications}  %
\newacronym{gspm}{GSpM}{Generalized spatial modulation}
\newacronym{gwp}{GWP}{Global Warming Potential}
\newacronym{harq}{HARQ}{hybrid automatic repeat request}
\newacronym{hat}{HAT}{hardware attached on top}
\newacronym{hcs}{HCS}{human-centric services}
\newacronym{hdf5}{HDF5}{Hierarchical Data Format version 5}
\newacronym{hfss}{HFSS}{High Frequency Simulator Software}
\newacronym{hmd}{HMD}{head-mounted display}
\newacronym{hpbm}{HPBM}{half power beam width}
\newacronym{HyMPRo}{HyMPRo}{Hybrid Multi-Path Routing algorithm}
\newacronym{i.i.d.}{i.i.d.}{independent and identically distributed}
\newacronym{i2c}{I2C}{Inter-Integrated Circuit}
\newacronym{iaq}{IAQ}{Indoor Air Quality}
\newacronym{ib}{IB}{in-band}
\newacronym{ibbc}{IBBC}{inter-band beam configuration}
\newacronym{ibo}{IBO}{input back-off}
\newacronym{ic}{IC}{integrated circuit}
\newacronym{iccs}{ICCS}{Ilmsens correlative channel sounder}
\newacronym{ici}{ICI}{intercarrier interference}
\newacronym{icnirp}{ICNIRP}{International Commission on Non-Ionizing Radiation Protection}
\newacronym{id}{ID}{information decoding}
\newacronym{idft}{IDFT}{inverse discrete Fourier transform}
\newacronym{idxm}{IDXM}{index modulation}
\newacronym{if}{IF}{intermediate-frequency}
\newacronym{ifft}{IFFT}{inverse fast-Fourier-transform}
\newacronym{iid}{i.i.d.}{independently and identically distributed}
\newacronym{iis}{IIS}{integrated information system}
\newacronym{im}{IM}{intermodulation}
\newacronym{imd}{IMD}{intermodulation distortion}
\newacronym{imu}{IMU}{inertial measurement unit}
\newacronym{inh}{InH}{indoor hotspot office}
\newacronym{io}{IO}{input/output}
\newacronym{ioe}{IoE}{Internet of Everything}
\newacronym{iot}{IoT}{Internet of Things}
\newacronym{ipt}{IPT}{inductive power transfer}
\newacronym{ipy}{IPY}{Interventions per Year}
\newacronym{iq}{IQ}{in-phase and quadrature}
\newacronym{iqi}{IQI}{IQ imbalance}
\newacronym{ir}{IR}{infrared}
\newacronym{isi}{ISI}{intersymbol interference}
\newacronym{ism}{ISM}{industrial, scientific and medical}
\newacronym{isp}{ISP}{internet service provider}
\newacronym{itu}{ITU}{International Telecommunication Union}
\newacronym{jesd}{JESD}{Joint Electron Devices Engineering Council}
\newacronym{jfet}{JFET}{junction field effect transistor}
\newacronym{kpi}{KPI}{key performance indicator}
\newacronym{ktofdm}{KT-DFT-s-OFDM}{known-tail-DFT-s-OFDM}
\newacronym{kvi}{KVI}{key value indicator}
\newacronym{larva}{LARVA}{LARge Virtual Array}
\newacronym{lca}{LCA}{life cycle assessment}
\newacronym{lco}{LCO}{lithium cobalt oxide}
\newacronym{ldo}{LDO}{Low-dropout voltage regulator}
\newacronym{ldpc}{LDPC}{low-density parity-check}
\newacronym{led}{LED}{Light Emitting Diode}
\newacronym{less}{LESS}{Low Energy Scheduler Solution}
\newacronym{lfp}{LFP}{lithium iron phosphate}
\newacronym{lib}{LIB}{Lithium-Ion Battery}
\newacronym{lic}{LIC}{lithium-ion capacitor}
\newacronym{lid}{LID}{Lithium Iron Disulfide}
\newacronym{lidar}{LiDAR}{light detection and ranging}
\newacronym{liion}{Li-ion}{lithium-ion}
\newacronym{lipo}{LiPo}{lithium polymer}
\newacronym{lis}{LIS}{large intelligent surface}
\newacronym{llh}{LLH}{log-likelihood}
\newacronym{lls}{LLS}{link-level simulation}
\newacronym{lmd}{LMD}{Lithium Manganese Dioxide}
\newacronym{lmmse}{LMMSE}{least minimum mean square error}
\newacronym{lmo}{LMO}{lithium ion manganese oxide}
\newacronym{lna}{LNA}{low-noise amplifier}
\newacronym{lo}{LO}{local oscillator}
\newacronym{lora}{LoRa}{long range}
\newacronym{lorawan}{LoRaWAN}{long-range wide-area network}
\newacronym{los}{LoS}{line-of-sight}
\newacronym{lp}{LP}{linear programming}
\newacronym{lpf}{LPF}{low-pass filter}
\newacronym{lpt}{LPT}{laser power transfer}
\newacronym{lpwa}{LPWA}{Low Power Wide Area}
\newacronym{lpwan}{LPWAN}{low-power wide-area network}
\newacronym{lpwans}{LPWANs}{Low-Power Wide-Area Networks}
\newacronym{lqi}{LQI}{link quality indicator}
\newacronym{lrelu}{LReLU}{leaky rectified linear unit}
\newacronym{lrt}{LRT}{likelihood-ratio test}
\newacronym{ls}{LS}{least squares}
\newacronym{lsa}{LSA}{large synthetic array}
\newacronym{lsf}{LSF}{large-scale fading}
\newacronym{lsfc}{LSFC}{large-scale fading component}
\newacronym{lstm}{LSTM}{Long Short-Term Memory}
\newacronym{ltc}{LTC}{lithium thionyl chloride}
\newacronym{lte}{LTE}{Long Term Evolution}
\newacronym{lti}{LTI}{linear time-invariant}
\newacronym{lto}{LTO}{lithium titanate}
\newacronym{lusta}{LUSTA 5G }{Logistique mUltimodale Sécuritaire Téléopérée \& Autonome 5G}
\newacronym{m2m}{M2M}{machine to machine}
\newacronym{mac}{MAC}{Medium Access Control}
\newacronym{mate}{MATE}{millimeter-wave MIMO testbed}
\newacronym{mc}{MC}{Monte Carlo}
\newacronym{mcl}{MCL}{Maximum Coupling Loss}
\newacronym{mcs}{MCS}{modulation and coding scheme}
\newacronym{mcu}{MCU}{microcontroller unit}
\newacronym{mec}{MEC}{multi-access edge computing}
\newacronym{mems}{MEMS}{micro-electromechanical systems}
\newacronym{mf}{MF}{matched filter}
\newacronym{mimo}{MIMO}{multiple-input multiple-output}
\newacronym{miso}{MISO}{multiple-input single-output}
\newacronym{ml}{ML}{machine learning}
\newacronym{mlp}{MLP}{multilayer perceptron}
\newacronym{mmic}{MMIC}{monolithic microwave integrated circuit}
\newacronym{mmimo}{mMIMO}{massive MIMO}
\newacronym{mmse}{MMSE}{minimum mean square error}
\newacronym{mmtc}{mMTC}{massive machine-typed communication}
\newacronym{mmwave}{mmWave}{millimeter wave}
\newacronym{mn}{MN}{matching network}
\newacronym{mosfet}{MOSFET}{metal-oxide semiconductor field effect transistor}
\newacronym{mpc}{MPC}{multipath component}
\newacronym{mppt}{MPPT}{maximum power point tracking}
\newacronym{mr}{MR}{maximum ratio}
\newacronym{mrc}{MRC}{maximum ratio combining}
\newacronym{mrc_em}{MRC}{maximum ratio combining}
\newacronym{mrc_EM}{MRC}{Magnetic Resonance Coupling}
\newacronym{mrt}{MRT}{maximum ratio transmission}
\newacronym{mse}{MSE}{mean square error}
\newacronym{msk}{MSK}{Minimum-Shift Keying}
\newacronym{mtc}{MTC}{Machine-Type Communication}
\newacronym{multi-rat}{Multi-RAT}{multiple radio access technology}
\newacronym{multirat}{Multi-RAT}{Multiple Radio Access Technology}
\newacronym{music}{MUSIC}{MUltiple SIgnal Classification}
\newacronym{navauwall}{NAVAUWALL}{AUtomated NAVigation in WALLonia}
\newacronym{nb}{NB}{narrowband}
\newacronym{nbiot}{NB-IoT}{narrowband IoT}
\newacronym{nca}{NCA}{nickel cobalt aluminum}
\newacronym{netcdf}{NetCDF}{Network Common Data Form}
\newacronym{nf}{NF}{noise figure}
\newacronym{nfc}{NFC}{near-field communication}
\newacronym{nfv}{NFV}{network function virtualization}
\newacronym{ngmn}{NGMN}{Next Generation Mobile Networks }
\newacronym{ni}{NI}{National Instruments}
\newacronym{nicd}{NiCd}{nikkel cadmium}
\newacronym{nimh}{NiMH}{nikkel metal hydride}
\newacronym{nlos}{NLoS}{non-line-of-sight}
\newacronym{nmc}{NMC}{nickel manganese cobalt}
\newacronym{nmos}{nMOS}{n-channel metal-oxide semiconductor}
\newacronym{nn}{NN}{neural network}
\newacronym{nnls}{NNLS}{non-negative least squares}
\newacronym{noma}{NOMA}{non-orthogonal multiple access}
\newacronym{np}{NP}{Neyman-Pearson}
\newacronym{npbch}{NPBCH}{Narrowband Physical Broadcast Channel}
\newacronym{npss}{NPSS}{Narrow Band Primay Synchronization Signal}
\newacronym{nr}{NR}{New Radio}
\newacronym{nrs}{NRS}{Narrow Band Reference Signal}
\newacronym{nsss}{NSSS}{Narrowband Secondary Synchronization Signal}
\newacronym{ntp}{NTP}{network time protocol}
\newacronym{oai}{OAI}{OpenAirInterface} %
\newacronym{obw}{OBW}{occupied bandwidth}
\newacronym{ofdm}{OFDM}{orthogonal frequency-division multiplexing}
\newacronym{ofdma}{OFDMA}{orthogonal frequency-division multiple access}
\newacronym{ofdmim}{OFDM-IM}{OFDM with index modulation}
\newacronym{olos}{OLoS}{obstructed-line-of-sight}
\newacronym{oma}{OMA}{orthogonal multiple access}
\newacronym{oob}{OOB}{out-of-band}
\newacronym{ook}{OOK}{on-off keying}
\newacronym{opbo}{OPBO}{output power backoff}
\newacronym{oran}{O-RAN}{open radio-access network}
\newacronym{os}{OS}{operating system}
\newacronym{ota}{OTA}{over-the-air}
\newacronym{otaa}{OTAA}{over-the-air authentication}
\newacronym{p1}{P1}{Phase 1}
\newacronym{p2}{P2}{Phase 2}
\newacronym{p2p}{P2P}{point-to-point}
\newacronym{pa}{PA}{power amplifier}
\newacronym{pae}{PAE}{power-added efficiency}
\newacronym{pam}{PAM}{pulse amplitude modulation}
\newacronym{pana}{PanA}{Panel A}
\newacronym{panb}{PanB}{Panel B}
\newacronym{papr}{PAPR}{peak-to-average power ratio}
\newacronym{pc}{PC}{pilot count}
\newacronym{pcb}{PCB}{printed circuit board}
\newacronym{pcg}{PCG}{power consumption gain}
\newacronym{pcie}{PCIe}{Peripheral Component Interconnect Express}
\newacronym{pcsi}{PCSI}{perfect channel state information}
\newacronym{pd}{PD}{powered device}
\newacronym{pdcch}{PDCCH}{physical downlink control channel}
\newacronym{pdf}{PDF}{probability density function}
\newacronym{pdp}{PDP}{power delay profile}
\newacronym{pdsch}{PDSCH}{physical downlink shared channel}
\newacronym{pe}{PE}{processing element}
\newacronym{peb}{PEB}{positioning error bound}
\newacronym{per}{PER}{packet error rate}
\newacronym{pet}{PET}{privacy enhancing technology}
\newacronym{pg}{PG}{path gain}
\newacronym{pgd}{PGD}{proximal gradient descent}
\newacronym{phy}{PHY}{physical}
\newacronym{pki}{PKI}{public key infrastucture}
\newacronym{pl}{PL}{path loss}
\newacronym{pla}{PLA}{physically large array}
\newacronym{pll}{PLL}{phase-locked loop}
\newacronym[plural=PMs,firstplural=person months (PMs)]{pm}{PM}{person month}
\newacronym{pmf}{PMF}{polymer microwave fiber}
\newacronym{pmu}{PMU}{power management unit}
\newacronym{pn}{PN}{pseudo-noise}
\newacronym{po}{PO}{phase offset}
\newacronym{poc}{PoC}{proof of concept}
\newacronym{poe}{PoE}{power-over-Ethernet}
\newacronym{pps}{1PPS}{pulse per second}
\newacronym{pr}{PR}{phase reversal}
\newacronym{prb}{PRB}{Physical Resource Block}
\newacronym{prbs}{PRBs}{Physical Resource Blocks}
\newacronym{prs}{PRS}{Peripheral Reflex System}
\newacronym{ps}{PS}{Processing System}
\newacronym{psd}{PSD}{power spectral density}
\newacronym{pse}{PSE}{power sourcing equipment}
\newacronym{psk}{PSK}{phase shift keying}
\newacronym{psm}{PSM}{power saving mode}
\newacronym{pss}{PSS}{primary synchronisation signal}
\newacronym{ptp}{PTP}{precision-time protocol}
\newacronym{ptrs}{PTRS}{Phase-Tracking Reference Signals}
\newacronym{ptw}{PTW}{paging time window}
\newacronym{pv}{PV}{photovoltaic}
\newacronym{pw}{PW}{planar wavefront}
\newacronym{pwm}{PWM}{pulse width modulation}
\newacronym{qam}{QAM}{quadrature amplitude modulation}
\newacronym{qos}{QoS}{quality-of-service}
\newacronym{qpsk}{QPSK}{quadrature phase-shift keying}
\newacronym{qrrls}{QR-RLS}{QR decomposition based recursive least squares}
\newacronym{quadriga}{QuaDRiGa}{QUAsi Deterministic RadIo channel GenerAtor}
\newacronym{ra}{RA}{Random Access}
\newacronym{ram}{RAM}{random-access memory}
\newacronym{ran}{RAN}{radio access network}
\newacronym{rar}{RAR}{Random Access Response}
\newacronym{rat}{RAT}{radio access technology}
\newacronym{rb}{Rb}{Rubidium}
\newacronym{rbs}{RBS}{radio base station}
\newacronym{rbw}{RBW}{resolution bandwidth}
\newacronym{rc}{RC}{raised-cosine}
\newacronym{rcs}{RCS}{radar cross section}
\newacronym{rdl}{RDL}{redistribution layer}
\newacronym{re}{RE}{radio element}
\newacronym{relu}{ReLU}{rectified linear unit}
\newacronym{rf}{RF}{radio frequency}
\newacronym{rfeh}{RFEH}{radio frequency energy harvesting}
\newacronym{rfic}{RFIC}{radio-frequency integrated circuit}
\newacronym{rfid}{RFID}{radio frequency identification}
\newacronym{rfpt}{RFPT}{radio frequency power transfer}
\newacronym{rfsoc}{RFSoC}{Radio Frequency System-on-Chip}
\newacronym{rir}{RIR}{room impulse response}
\newacronym{ris}{RIS}{reflective intelligent surface}%
\newacronym{rllmtc}{RLLMTC}{reliable low latency machine type communication}
\newacronym{rls}{RLS}{recursive least squares}
\newacronym{rms}{RMS}{root-mean-square}
\newacronym{rmse}{RMSE}{root-mean-square error}
\newacronym{rmt}{RMT}{random matrix theory}
\newacronym{rof}{RoF}{radio-over-fiber}
\newacronym{ros}{ROS}{robot operating system}
\newacronym{rpi}{RPi}{Raspberry Pi}
\newacronym{rrc}{RRC}{Radio Resource Connection}
\newacronym{rreq}{RREQ}{route request packet}
\newacronym{rsrp}{RSRP}{Reference Signals Received Power}
\newacronym{rsrq}{RSRQ}{Reference Signal Received Quality}
\newacronym{rss}{RSS}{received signal strength}
\newacronym{rssi}{RSSI}{received signal strength indicator}
\newacronym{rtc}{RTC}{real time clock}
\newacronym{rtf}{RTF}{reader talks first}
\newacronym{rtk}{RTK}{real time kinematics}
\newacronym{rts}{RTS}{ray tracing simulator}
\newacronym{ru}{RU}{Radio Unit}
\newacronym{rv}{RV}{random variable}
\newacronym{rw}{RW}{RadioWeaves}
\newacronym{rx}{RX}{receiver}
\newacronym{rzf}{RZF}{regularized zero forcing}
\newacronym{s-parameter}{S-parameter}{scattering parameter}
\newacronym{sa}{SA}{synchronization anchor}
\newacronym{sa5}{SA}{Stand Alone}
\newacronym{sar}{SAR}{specific absorption rate}
\newacronym{sbl}{SBL}{sparse Bayesian learning}
\newacronym{sc}{SC}{single carrier}
\newacronym{scfde}{SC-FDE}{single-carrier modulation with frequency-domain-equalization}
\newacronym{scfdma}{SCFDMA}{single-carrier frequency division multiple access}
\newacronym{scs}{SCS}{sub-carrier spacing}
\newacronym{sdg}{SDG}{Sustainable Development Goal}
\newacronym{sdm}{SDM}{sigma-delta modulator}
\newacronym{sdma}{SDMA}{spatial-division multiple access}
\newacronym{sdn}{SDN}{software-defined network}
\newacronym{sdof}{SDoF}{sigma-delta over fiber}
\newacronym{sdr}{SDR}{software-defined radio}
\newacronym{se}{SE}{spectral efficiency}
\newacronym{sei}{SEI}{specific emitter identification}
\newacronym{ser}{SER}{symbol-error rate}
\newacronym{sf}{SF}{spreading factor}
\newacronym{sfn}{SFN}{single frequency network}
\newacronym{sfp}{SFP}{small form-factor pluggable}
\newacronym{sha}{SHA}{Secure Hash Algorithm}
\newacronym{SigMF}{SigMF}{Signal Metadata Format}
\newacronym{simo}{SIMO}{single-input multiple-output}
\newacronym{sinr}{SINR}{signal-to-interference-plus-noise ratio}
\newacronym{siso}{SISO}{single-input single-output}
\newacronym{slam}{SLAM}{simultaneous localization and mapping}
\newacronym{slc}{SLC}{spatial leakage suppression}
\newacronym{slerp}{SLERP}{spherical linear interpolation}
\newacronym{sma}{SMA}{SubMiniature version A}
\newacronym{smc}{SMC}{specular multipath component}
\newacronym{smps}{SMPS}{switched mode power supply}
\newacronym{sndr}{SNDR}{signal-to-noise-and-distortion ratio}
\newacronym{snidr}{SNIDR}{signal-to-noise-and-interference-and-distortion ratio}
\newacronym{snir}{SNIR}{signal-to-interference-plus-noise ratio}
\newacronym{snr}{SNR}{signal-to-noise ratio}
\newacronym{soc}{SoC}{state of charge}
\newacronym{SoC}{SOC}{System on Chip}
\newacronym{sp}{SP}{service point}
\newacronym{spdt}{SPDT}{single pole double throw}
\newacronym{spi}{SPI}{Serial Peripheral Interface}
\newacronym{spst}{SPST}{single pole single throw}
\newacronym{sram}{SRAM}{static random-access memory}
\newacronym{srd}{SRD}{short-range device}
\newacronym{srls}{SRLS}{standard recursive least squares}
\newacronym{srs}{SRS}{Sounding Reference Signal}
\newacronym{ssb}{SSB}{synchronisation signal block}
\newacronym{ssd}{SSD}{solid state drive}
\newacronym{ssq}{SSQ}{simulator sickness questionnaire}
\newacronym{steam}{STEAM}{science, technology, engineering, the arts, and mathematics}
\newacronym{svd}{SVD}{singular value decomposition}
\newacronym{sw}{SW}{spherical wavefront}
\newacronym{swipt}{SWIPT}{simultaneous wireless information and power transfer}
\newacronym{synce}{SyncE}{Synchronous Ethernet}
\newacronym{ta}{TA}{timing advance}
\newacronym{tau}{TAU}{tracking area update}
\newacronym{tcer}{TCER}{transported to consumed energy ratio}
\newacronym{tcp}{TCP}{Transmission Control Protocol}
\newacronym{tcxo}{TCXO}{temperature compensated crystal oscillator}
\newacronym{tdce}{TD-CE}{time-domain compression and expansion}
\newacronym{tdd}{TDD}{time division duplexing}
\newacronym{tdma}{TDMA}{time division-multiple access}
\newacronym{tdoa}{TDOA}{time-difference-of-arrival}
\newacronym{thz}{THz}{Terahertz}
\newacronym{to}{TO}{timing offset}
\newacronym{toa}{TOA}{time-of-arrival}
\newacronym{tof}{ToF}{time-of-flight}
\newacronym{tosm}{TOSM}{through-open-short-match}
\newacronym{tpms}{TPMS}{Tire-Pressure Monitoring System}
\newacronym{trl}{TRL}{technology readyness level}
\newacronym{trp}{TRP}{Transmission Reception Point}
\newacronym{tsn}{TSN}{time-sensitive networking}
\newacronym{ttf}{TTF}{tag talks first}
\newacronym{ttff}{TTFF}{Time To First Fix}
\newacronym{tti}{TTI}{transmission time interval}
\newacronym{ttm}{TTM}{time to market}
\newacronym{ttn}{TTN}{The Things Network}
\newacronym{tx}{TX}{transmitter}
\newacronym{uart}{UART}{Universal Asynchronous Receiver/Transmitter}
\newacronym{uav}{UAV}{unmanned aerial vehicle}
\newacronym{uc}{UC}{use case}
\newacronym{ucie}{UCIe}{Universal Chiplet Interconnect Express}
\newacronym{udp}{UDP}{User Datagram Protocol}
\newacronym{ue}{UE}{user equipment}
\newacronym{ugv}{UGV}{unmanned ground vehicle}
\newacronym{uhd}{UHD}{USRP hardware driver}
\newacronym{uhf}{UHF}{ultra-high frequency}
\newacronym{ul}{UL}{uplink}
\newacronym{ula}{ULA}{uniform linear array}
\newacronym{uMIMO}{$\mu$-MIMO}{ultra-massive Multiple-Input Multiple-Output}
\newacronym{ummtc}{umMTC}{ultra massive machine type communication}
\newacronym{un}{UN}{United Nations}
\newacronym{unow}{uNOW}{unified non-orthogonal waveform}
\newacronym{upa}{UPA}{uniform planar array}
\newacronym{ura}{URA}{uniform rectangular array}
\newacronym{urllc}{URLLC}{ultra-reliable low-latency communications}
\newacronym{usrp}{USRP}{universal software radio peripheral}
\newacronym{uv}{UV}{unmanned vehicle}
\newacronym{uw}{UW}{unique word}
\newacronym{uwb}{UWB}{ultrawideband}
\newacronym{uwofdm}{UW-DFT-s-OFDM}{unique word DFT-s-OFDM}
\newacronym{v2v}{V2V}{vehicle-to-vehicle}
\newacronym{vco}{VCO}{voltage-controlled oscillator}
\newacronym{vep}{VEP}{virtual edge platform}
\newacronym{vlc}{VLC}{visible light communication}
\newacronym{vlp}{VLP}{visible light positioning}
\newacronym{vna}{VNA}{vector network analyzer}
\newacronym{voc}{VOC}{Voltatile Organic Compound}
\newacronym{votable}{VOTable}{Virtual Observatory Table}
\newacronym{vr}{VR}{virtual reality}
\newacronym{vswr}{VSWR}{voltage standing wave ratio}
\newacronym{vuca}{VUCA}{volatile, uncertain, complex and ambiguous}
\newacronym{wb}{WB}{wideband}
\newacronym{wban}{WBAN}{wireless body area network}
\newacronym{wd}{WD}{Wireless distance}
\newacronym{wimax}{WiMAX}{Worldwide Interoperability for Microwave Access}
\newacronym{wlan}{WLAN}{wireless LAN}
\newacronym[plural=WPs,firstplural=work packages (WPs)]{wp}{WP}{work package}
\newacronym{wpt}{WPT}{wireless power transfer}
\newacronym{wr}{WR}{White Rabbit}
\newacronym{wrsn}{WRSN}{wireless rechargeable sensor network}
\newacronym{wsn}{WSN}{wireless sensor network}
\newacronym{xets}{XETS}{cross exponentially tapered slot}
\newacronym{xlmimo}{XL-MIMO}{extremely large-scale MIMO}
\newacronym{xr}{XR}{extended reality}
\newacronym{z3ro}{Z3RO}{zero third-order distortion}
\newacronym{zf}{ZF}{zero-forcing}
\newacronym{zmcscg}{ZMCSCG}{zero mean circularly symmetric complex Gaussian}
\newacronym{zmq}{ZMQ}{ZeroMQ}
\newacronym{ztofdm}{ZT-DFT-s-OFDM}{zero-tail DFT-s-OFDM}
\newacronym{llr}{LLR}{log-likelihood ratio}           %

\input{auto_names}                  %

\usepackage{orcidlink}

\usepackage[backend=biber,style=ieee]{biblatex}
\AtBeginBibliography{\footnotesize}

\def\BibTeX{{\rm B\kern-.05em{\sc i\kern-.025em b}\kern-.08em
    T\kern-.1667em\lower.7ex\hbox{E}\kern-.125emX}}

\definecolor{color1}{HTML}{1b9e77}
\definecolor{color2}{HTML}{d95f02}
\definecolor{color3}{HTML}{7570b3}
\definecolor{color4}{HTML}{e7298a}
\definecolor{color5}{HTML}{66a61e}
\definecolor{color6}{HTML}{e6ab02}
\definecolor{color7}{HTML}{a6761d}
\definecolor{color8}{HTML}{666666}

\colorlet{AMBgreen}{color1}
\colorlet{AMBdarkgreen}{color1}
\colorlet{AMBlightgreen}{color1}
\definecolor{darkgray176}{RGB}{176,176,176}
\usepackage{tikz}
\usepackage{pgfplots}   %
\pgfplotsset{compat=newest}
\usetikzlibrary{plotmarks}
\usetikzlibrary{arrows.meta}
\usetikzlibrary{arrows}
\usetikzlibrary{calc,fadings,decorations.pathreplacing}
\usetikzlibrary{fit,backgrounds}  %

\usetikzlibrary{shapes,arrows,fit,positioning}
\usetikzlibrary{shapes}
\usetikzlibrary{spy}
\usetikzlibrary{circuits}
\usetikzlibrary{arrows}

\pgfplotsset{
    every axis/.append style={
        title style={draw=none},
        label style={font=\small},
        legend style={
            fill opacity=0.8,
            nodes={scale=0.8, transform shape}, {draw=none}
        },
        tick align=outside,
        tick pos=left,
        x grid style={darkgray176},
        xtick style={color=black},
        y grid style={darkgray176},
        ytick style={color=black},
        grid=both,
    },
    every axis plot/.append style={
        line width=2.0pt,
    },
}

\tikzset{%
  >=latex,
  inner sep=0pt,%
  outer sep=2pt,%
  mark coordinate/.style={inner sep=0pt,outer sep=0pt,minimum size=3pt,
  fill=black,circle}%
}

\usepackage[font=small]{caption}
\usepackage{subcaption}

\usepackage{fontawesome}
\usetikzlibrary {patterns,patterns.meta}
\usetikzlibrary{arrows.meta,positioning,calc}
\usetikzlibrary{intersections}
\usepackage{placeins}

\defineauthors[false]{gilles, tijl, liesbet}

\begin{document}
\title{Towards Energy-Neutral IoT Sensors: Low-Cost Chirp-Based Backscattering Node Using a COTS Microcontroller and Multi-Load Front-End
\thanks{This work was supported by the AMBIENT-6G and SUSTAIN-6G project, which received funding from the Smart Networks and Services Joint Undertaking (SNS JU) under the European Union's Horizon Europe research and innovation programme under Grant Agreement No. 101192113 and 101191936, respectively.}
\thanks{This work has been accepted at IEEE Sensors 2026.} %
}

\author{
\IEEEauthorblockN{%
Tijl Schepens\,\orcidlink{0009-0007-7177-4141}\IEEEauthorrefmark{1},
Liesbet Van der Perre\,\orcidlink{0000-0002-9158-9628}\IEEEauthorrefmark{1},
Gilles Callebaut\,\orcidlink{0000-0003-2413-986X}\IEEEauthorrefmark{1}
}
\IEEEauthorblockA{\IEEEauthorrefmark{1}Department of Electrical Engineering, KU Leuven, Belgium}
}

\maketitle

\begin{abstract}
Backscatter communication has proven to be a key enabler for energy-neutral sensor nodes in the \gls{iot}. It allows ultra-low-power nodes to transmit sensor data by reflecting incident \gls{rf} signals. Practical large-scale deployment requires hardware solutions that are both cost-efficient and widely available. 
This paper investigates the practical limits of implementing backscatter transmitters exclusively using \gls{cots} components. We compare a two-load and multi-load architecture in terms of hardware requirements, complexity and spectral efficiency highlighting the most important trade-offs. A multi-load backscatter architecture is developed using a resource constrained low-power microcontroller, levering \gls{dma} to achieve high switching speeds while minimizing processing overhead. 
Furthermore, we present a layout agnostic load design methodology based on scattering parameters, allowing precise control of the reflection coefficients. A prototype implementation demonstrates a successful backscatter transmission of a single-sideband chirp with reduced harmonic distortion. Measurement results show suppression of the backscattered carrier and the second sideband by 10\,dBm. This demonstrates that low-cost sensor nodes can be built using \gls{cots} components enabling remote monitoring at extended lifetimes. %
\end{abstract}

\begin{IEEEkeywords}
Internet of Things, Energy-neutral Devices, Direct Memory Access, Backscatter communication
\end{IEEEkeywords}

\section{Introduction}
Sensor nodes balance energy harvesting and consumption attempting to reach energy-neutral operation. Backscattering emerged as a key technology due to its low power consumption. It enables \gls{iot} sensor nodes to perform remote monitoring without spending a significant energy budget on wireless transmissions. However, mass deployments require low-cost hardware solutions combined with a minimal power budget. This conflicts with many research solutions relying on \glspl{fpga} \cite{wangAllSparkEnablingLongRange2022,liXORLoRaLoRaBackscatter2020,liNovelLoadFreeSSB2025,tallaLoRaBackscatterEnabling2017}. Others simplify the \gls{rf} front-end to a 2-load modulation scheme to operate on microcontrollers sacrificing spectral efficiency \cite{renAeroEchoAgriculturalLowpower2025,tangSelfSustainableLongRangeBackscattering2021,tangPrototypeImplementationExperimental2025}. While monolithic integration forms a solution for \gls{fpga} based solutions to reach low-cost, \gls{cots} have to bridge the gap and are the only solution for lower volumes. This paper explores the limits of implementing backscattering relying solely on \gls{cots} components. The explored solution is fully open and hosted on the \href{https://github.com/AMBIENT-6G/END-Platform}{Ambient-6G END Platform repository}\footnote{\url{https://github.com/AMBIENT-6G/END-Platform}}. It aims at deployment readiness by focusing on availability and cost optimization.
\begin{figure}
    \centering
    \includegraphics[width=0.8\columnwidth]{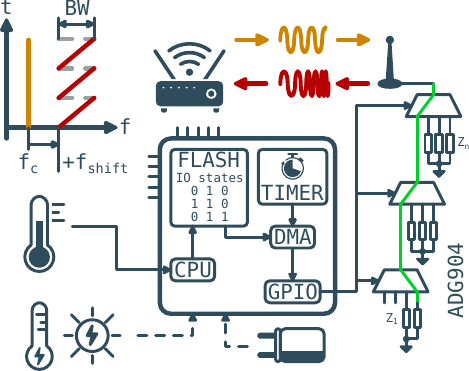}
    \caption{A low-power \gls{mcu} based sensor node backscattering a chirp by controlling its \glspl{gpio} through the \gls{dma} peripheral, connecting the right load to the antenna with a multiplexer.}
    \label{fig:mcu-dma-backscatter}
\end{figure}

\section{Hardware architectures}
The most important backscattering hardware architectures~\cite{schepens2026long} are 2-load \cite{tangSelfSustainableLongRangeBackscattering2021}, multi-load \cite{tallaLoRaBackscatterEnabling2017} and IQ-modulator \cite{beloIQImpedanceModulator2019} approaches. The component controlling the loads needs to be low-power and low-cost. This results in a microcontroller becoming the brain of the solution. Two load architectures form the basis of many experiments as they lower the \gls{mcu}'s requirements. Here a simple IO switching between high and low directly translates into a single tone being backscattered. Switching between two switching speeds or linearly changing the switching speed over time results in 2-\gls{fsk} \cite{varshneyLoReaBackscatterArchitecture2017} or chirp based modulation \cite{tangPrototypeImplementationExperimental2025} respectively. The base switching speed can easily be chosen sufficiently high to push the backscattered signals out-of-band avoiding complex carrier cancellation techniques and boosting range. The main downside comes from the reduced spectral efficiency due to the square wave like modulation. It creates significant harmonics which can cause disturbance in adjacent communication channels.

Multi-load solutions solve this by gradually changing the phase of the backscattered signal. To generate the same backscatter signal as a 2-load solution, a $n$-load system needs to cycle $\frac{n}{2}$ times faster between the loads. More complex modulations easily require significantly higher switching speeds because of this. For example a chirp with a bandwidth $BW$ of \SI{125}{\kilo\hertz} and shifted \SI{125}{\kilo\hertz} ($f_\text{shift}$) away from the carrier results in a maximum cycling frequency of $2(f_\text{shift} + BW) = \SI{500}{\kilo\hertz}$ with 2 loads, which becomes $\frac{n}{2} \cdot f_{max} = \frac{8}{2} \cdot \SI{500}{\kilo\hertz} = \SI{2}{\mega\hertz}$ for eight loads.

IQ-modulator approaches utilize an \gls{rf} transistor to continuously change the backscatter impedance by sweeping the gate voltage. By using two transistors where one signal branch gets shifted by \SI{90}{\degree} results in the quadrature operation. The transistors are either driven by a \gls{dac} or directly by \glspl{gpio}. Either the serial interface controlling the \gls{dac} needs to be fast enough or the \glspl{gpio} need sufficient oversampling \cite{liNovelLoadFreeSSB2025} to generate proper modulated waveforms.

\section{Multi-load front-end design}
The design starts with the selection of a low-power microcontroller. Here the STM32U031 from ST Microelectronics and the MSPM0C1104 from Texas Instruments are considered.

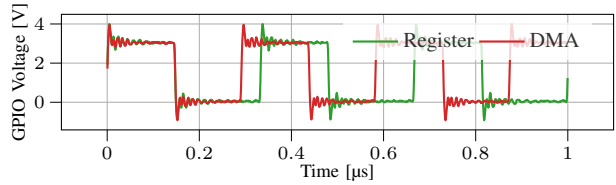
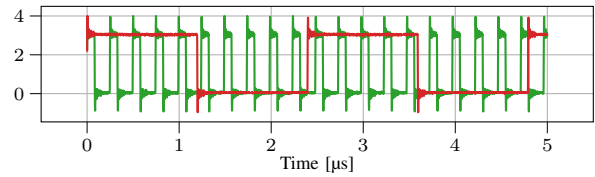
\begin{figure*}[htbp]
    \centering
    \begin{subfigure}[t]{0.49\textwidth}
        \centering
        \begin{tikzpicture}
            \begin{axis}[
                height=0.35\linewidth,
                width=\linewidth,
                xlabel={Time [µs]},
                ylabel={GPIO Voltage [V]},
                grid=major,
                legend pos=north east,
                legend columns=2,
                scaled x ticks=false,
                cycle list shift=2,
                        label style={font=\scriptsize},
        ticklabel style={font=\scriptsize}
            ]
            \addplot+[line width=0.8pt] table[
                col sep=comma,
                x expr=\thisrow{time}*1e6,
                y=CH1
            ] {figures/stm32-gpio-toggle-dec.csv};
            \addlegendentry{Register}

            \addplot+[line width=0.8pt] table[
                col sep=comma,
                x expr=\thisrow{time}*1e6,
                y=CH1
            ] {figures/stm32-dma-toggle-dec.csv};
            \addlegendentry{DMA}
            \end{axis}
        \end{tikzpicture}
        \caption{STM32}
        \label{fig:gpio-switching-speeds-stm32}
    \end{subfigure}%
    \begin{subfigure}[t]{0.49\textwidth}
        \centering
        \begin{tikzpicture}
            \begin{axis}[
                height=0.35\linewidth,
                width=\linewidth,
                xlabel={Time [µs]},
                grid=major,
                legend pos=north east,
                legend columns=2,
                scaled x ticks=false,
                cycle list shift=2,
                        label style={font=\scriptsize},
        ticklabel style={font=\scriptsize}
            ]
            \addplot+[line width=0.8pt] table[
                col sep=comma,
                x expr=\thisrow{time}*1e6,
                y=CH1
            ] {figures/mspm0-gpio-toggle-dec.csv};

            \addplot+[line width=0.8pt] table[
                col sep=comma,
                x expr=\thisrow{time}*1e6,
                y=CH1
            ] {figures/mspm0-dma-toggle-dec.csv};
            \end{axis}
        \end{tikzpicture}
        \caption{MSPM0}
        \label{fig:gpio-switching-speeds-mspm0}
    \end{subfigure}

    \caption{Comparison of the GPIO switching speeds achieved on a STM32U031 and a MSPM0C1104 \gls{mcu} using direct register access or the \gls{dma} peripheral triggered by a timer. Note that time scales differ.}
    \label{fig:gpio-switching-speeds}
\end{figure*}

Multi-load designs utilize a multiplexer between the loads and the antenna. The microcontroller decides which load gets connected to the antenna with appropriate speed. The switching speed depends on the microcontroller being used and on how they are controlled. Direct register access yields high switching speeds of \SI{3}{\mega\hertz} for the STM32U031 and \SI{4}{\mega\hertz} for the MSPM0C1104 respectively (\cref{fig:gpio-switching-speeds}). However it takes excessive \gls{cpu} time and the speed drops immediately when the \gls{cpu} handles more complex logic. The solution lies in the use of the \gls{dma} peripheral, offloading the \gls{cpu} and maintaining maximum switching speeds. The STM32U031 achieves an even slightly higher switching speed of \SI{3.4}{\mega\hertz} using this approach. The MSPM0C1104 is unable to go above \SI{400}{\kilo\hertz} because of the time required to re-start the \gls{dma} controller after a transfer. The STM32U031 operates in circular mode constantly resetting its memory address counter in hardware saving valuable \gls{cpu} time.

\section{Load design}
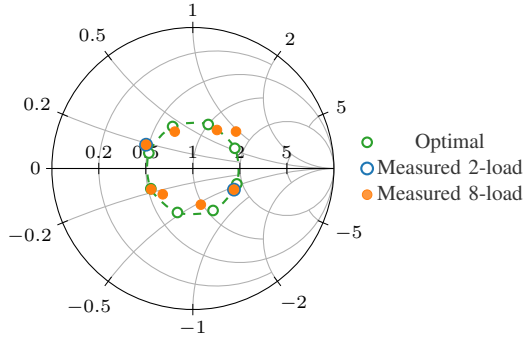
\begin{figure}[htbp]
    \centering
    \begin{tikzpicture}
        \begin{smithchart}[
            height=0.6\columnwidth,
            width=0.6\columnwidth,
            legend style={
            at={(1.05,0.5)},
            anchor=west,
            draw=none,
            fill=none,
            align=left,
        },
        legend columns=1,
        label style={font=\scriptsize},
        ticklabel style={font=\scriptsize}
        ]
          \draw[mplgreen, dashed, thick]
            plot[smooth cycle] coordinates {
              (0.5226,-0.1682)
              (0.6712,-0.4698)
              (1.0832,-0.7238)
              (1.8348,-0.4496)
              (1.7338,0.5578)
              (1,0.7)
              (0.6382,0.4266)
              (0.5142,0.126)
            };
        
        \addplot+[only marks, mark size=1.8pt,mplgreen, line width=0.8pt,
            mark options={
            fill=white,         %
            draw=mplgreen,      %
            draw opacity=1     %
        }] coordinates {
            (1.8348,-0.4496) (1.7338,0.5578) (0.6712,-0.4698) (1.0832,-0.7238) (0.5142,0.126) (0.6382,0.4266) (0.5226,-0.1682) (1,0.7)
        };
        \addlegendentry{Optimal}

        \addplot+[
        only marks, 
        mark size=2.2pt, 
        line width=0.8pt,
        draw=mplblue,
        fill=white,
        ] coordinates {
            (0.4786,0.1874) (1.702,-0.5714)
        };
        \addlegendentry{Measured 2-load}
        
         \addplot+[only marks, mark size=1.8pt,mplorange,line width=0pt] coordinates {
            (1.702,-0.5714) (1.5226,0.952) (0.612,-0.243) (0.9746,-0.5366) 
            (0.4786,0.1874) (0.6828,0.39) (0.5236,-0.1744) (1.1766,0.7168)
        };
        \addlegendentry{Measured 8-load}

        \end{smithchart}
    \end{tikzpicture}
    \caption{The theoretical impedance points in green lying on a circle and the practically measured points in orange and blue.}\label{fig:antenna-impedance-smith}
\end{figure}

The load impedances are selected such that their reflection coefficients lie on a constant \gls{vswr} circle in the Smith chart (\cref{fig:antenna-impedance-smith}). Cycling through the loads around this circle rotates the phase of the reflected wave. The resulting frequency shift is equal to the cycling rate. Because this phase rotation is smoother than switching between two loads, the backscattered signal is concentrated in one sideband and contains fewer harmonics~\cite{tallaLoRaBackscatterEnabling2017}.

This ideal load placement assumes that every load is connected to the antenna through an identical path. In practice, the \gls{pcb} traces and \gls{rf} switches between the antenna and each load add phase shift and attenuation. If all paths have equal length, this effect is the same for every load and only rotates the complete constant \gls{vswr} circle. In our design the paths have different lengths, so each load experiences a different phase shift and attenuation. Therefore, each load path (green path indicated in \cref{fig:mcu-dma-backscatter}) was treated as a two-port network and characterized with a \gls{vna} before selecting the final load values.

Each measured load path is described by four S-parameters. With port 1 at the antenna side and port 2 at the selected load, the incident and reflected waves are related by
\begin{align}
    b_1 &= S_{11} a_1 + S_{12} a_2 \label{eq:b1}\\
    b_2 &= S_{21} a_1 + S_{22} a_2 \label{eq:b2}
\end{align}
where $a_i$ denotes the wave incident on port $i$ and $b_i$ denotes the wave leaving port $i$.

The load value is chosen at port 2, but the target reflection coefficient is defined at the antenna port. Therefore, the load reflection coefficient $\Gamma_\text{L}$ has to be converted to the input reflection coefficient $\Gamma_\text{in}$. Using the wave convention above, these reflection coefficients are
\begin{align}
    \Gamma = \frac{\text{reflected wave}}{\text{incident wave}}\\
    \Gamma_{in} = \frac{b_1}{a_1}, \Gamma_L = \frac{a_2}{b_2}
\end{align}

Using $a_2 = \Gamma_L b_2$, \cref{eq:b2} can be rearranged to express $b_2$ as
\begin{align}
    a_2 &= \Gamma_L b_2\\
    b_2 &= \frac{S_{21} a_1}{1 - S_{22} \Gamma_L} \label{eq:b2gammal}
\end{align}

Substituting this result into \cref{eq:b1} gives
\begin{align}
    b_1 &= S_{11} a_1 + S_{12} \Gamma_L \frac{S_{21} a_1}{1 - S_{22} \Gamma_L}
\end{align}

Solving for $\Gamma_{in}$, and then inverting the relation for $\Gamma_{L}$, yields
\begin{align}
    \Gamma_{in} &= S_{11} + \frac{S_{12} S_{21} \Gamma_L}{1 - S_{22} \Gamma_L}\label{eq:gintogl}\\
    \Gamma_L &= \frac{\Gamma_{in} - S_{11}}{S_{12} S_{21} + S_{22}(\Gamma_{in} - S_{11})}
\end{align}
These are the standard two-port transformations from \cite{pozarMicrowaveEngineering2011}. They allow a desired $\Gamma_{in}$ to be chosen on the constant \gls{vswr} circle and translated into the required $\Gamma_{L}$ for each physical load path. After selecting the loads, the resulting input reflections were measured at the antenna port. The measured points are shown in \cref{fig:antenna-impedance-smith} and closely follow the theoretical points calculated with a Python script. The remaining deviations are attributed to parasitics and to a \gls{pcb} layout that was not optimized for equal load paths.

\subsection{Measurements}
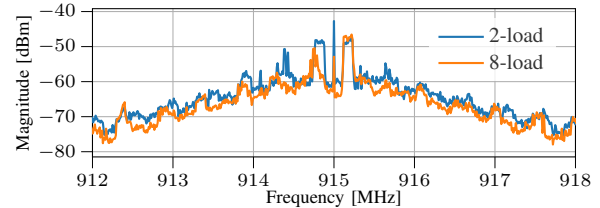
\begin{figure}[htbp]
    \centering
        \begin{tikzpicture}
            \begin{axis}[
                height=0.4\columnwidth,
                width=0.9\columnwidth,
                xlabel={Frequency [MHz]},
                ylabel={Magnitude [dBm]},
                grid=major,
                xmax=918,
                xmin=912,
                legend pos=north east,
                        label style={font=\scriptsize},
        ticklabel style={font=\scriptsize}
            ]
    
            \addplot+[line width=0.8pt] table[col sep=comma, x expr=\thisrow{Frequency}/1e6, y expr=\thisrow{PORT1_Magnitude}] {figures/rfvt-1/2load-chirp.csv};
            \addlegendentry{2-load}

            \addplot+[line width=0.8pt] table[col sep=comma, x expr=\thisrow{Frequency}/1e6, y expr=\thisrow{PORT1_Magnitude}] {figures/rfvt-1/8load-chirp-1.csv};
            \addlegendentry{8-load}

            \end{axis}
        \end{tikzpicture}

    \caption{The 8-load architecture produces a single sideband chirp while suppressing the backscattered carrier by \SI{10}{\dBm} and lowering the 3\textsuperscript{d} and 5\textsuperscript{th} harmonic. Measured using a LibreVNA and averaged over 10 measurements.}\label{fig:multi-load-backscatter}
\end{figure}

\begin{figure}[htbp]
    \centering
    \begin{tikzpicture}
\begin{axis}[
    width=0.65\columnwidth,
    height=0.42\columnwidth,
    scale only axis,
    xmin=915.300292969,
    xmax=915.549316406,
    ymin=0.000000000,
    ymax=10.137600000,
    xlabel={Frequency [MHz]},
    ylabel={Time [ms]},
    enlargelimits=false,
    axis on top,
    colorbar,
    point meta min=-55,
    point meta max=0,
    colormap={waterfallviridis}{
        rgb255=(68,1,84)
        rgb255=(59,82,139)
        rgb255=(33,145,140)
        rgb255=(94,201,98)
        rgb255=(253,231,37)
    },
    colorbar style={ylabel={Relative magnitude [dB]}},
            label style={font=\scriptsize},
        ticklabel style={font=\scriptsize}
]
\addplot graphics[
    xmin=915.300292969,
    xmax=915.549316406,
    ymin=0.000000000,
    ymax=10.137600000
] {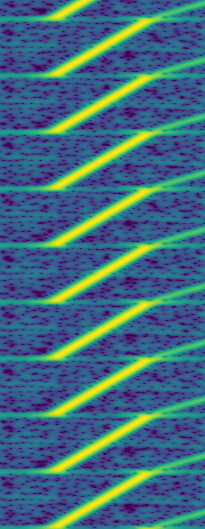};
\addplot[
    draw=none,
    mark=none,
    forget plot,
    point meta=explicit
] coordinates {
    (915.300292969,0.000000000) [-55]
    (915.549316406,10.137600000) [0]
};
\end{axis}
\end{tikzpicture}
    \caption{A waterfall plot shows the backscatter device constantly sending a single up-chirp.}\label{fig:waterfall-plot}
\end{figure}

The spectrally efficient \gls{dma} solution is tested on the STM32U031 microcontroller running at \SI{48}{\mega\hertz}. Three ADG904 switches form the 8 load multiplexer network connected to the antenna. The microcontroller IO pins are directly connected to the multiplexer switches. For the transmitter a LibreVNA is used transmitting a constant carrier at $f_\text{c} = \SI{915}{\mega\hertz}$ with an output power of \SI{-10}{\dBm}. A second LibreVNA measures the backscattered signal in spectrum analyzer mode. To cancel the carrier and avoid including the path loss in the measurements a circulator is used (TH2528XS-1).
Measurements show that using 8 loads results in a single-sideband signal with more energy concentrated in it (\cref{fig:multi-load-backscatter}). The backscattered carrier gets suppressed by more than \SI{10}{\dBm}. In practice this increases the achievable range of the tag and reduces interference in adjacent bands. A waterfall plot (\cref{fig:waterfall-plot}) confirms that the backscattered signal is a proper linear chirp between \num{125}~and~\SI{250}{\kilo\hertz} ($f_\text{shift} = \SI{125}{\kilo\hertz}, BW = \SI{125}{\kilo\hertz}$). The residual mirror image in the 8 load scenario comes from the imperfect load spacing which can be solved by further load tuning. The \gls{rf} front-end consumes \SI{81.14}{\nano\watt} while idle and \SI{3.63}{\micro\watt} during a backscatter transmission.

\glsresetall
\section{Conclusion}
Our solution demonstrates that \gls{cots} solutions relying on microcontrollers can build spectral efficient backscatter solutions. While the multi-load hardware architectures produce single sideband signals and reduce harmonics they require higher IO switching speeds. Relying on the \gls{dma} peripheral and storing the full waveform image in flash memory overcomes this hurdle. Even a small microcontroller with a Cortex-M0+ core operating at \SI{48}{\mega\hertz} can backscatter chirps shifted away from the carrier. The \gls{dma} needs support for repeated transfers and a high enough trigger rate from a timer. While the microcontroller and backscatter loads are low-cost in high volumes the \gls{rf} switch network poses a significant cost. The switches come in threefold to support eight loads already costing 7.5 euro\footnote{Price at supplier for more than \num{10000} pieces.}. \Gls{spdt} switches require 7 parts to reach 8 loads but come at a much lower price. The FSA3051TMX posses a bandwidth of \SI{1}{\giga\hertz} and only costing 1.4 euros in total.

Other hardware architectures such as the IQ-modulator also allow spectral efficient transmissions. Further research can focus on these other architectures to find out if simple microcontroller solutions can reach the required sample rates.

\printbibliography%

\end{document}